\documentclass[sigconf]{acmart}

\usepackage{amsmath}
\usepackage[ruled,vlined]{algorithm2e}
\usepackage[T1]{fontenc}
\usepackage{graphicx}
\usepackage{textcomp}
\usepackage{tikz}
\usetikzlibrary{shapes,arrows}
\usepackage{hyperref}
\usepackage{textcomp}
\usepackage{color}
\usepackage[utf8]{inputenc}
\usepackage{fancyvrb}
\usepackage{xcolor}
\usepackage{colortbl}
\usepackage{hhline}
\usepackage{lscape}
\usepackage{upquote}
\usepackage{array}
\usepackage{adjustbox}
\usepackage{rotating}
\usepackage{stmaryrd}
\usepackage{pifont}
\usepackage{multirow}
\usepackage{wrapfig}
\usepackage{enumitem}
\usepackage{balance}
\usepackage[clock]{ifsym}
\usepackage{flushend}
\usepackage{forest}

\usepackage[utf8]{inputenc}
\usepackage{lineno,hyperref}
\usepackage{colortbl}
\usepackage{rotating}
\usepackage{pgfplots,pgfplotstable}
\usepackage{scalefnt}

\usetikzlibrary{calc}

\newcolumntype{L}[1]{>{\raggedright\arraybackslash}m{#1}}
\newcolumntype{C}[1]{>{\centering\arraybackslash}m{#1}}
\newcolumntype{R}[1]{>{\raggedleft\arraybackslash}m{#1}}

\fvset{commandchars=\\\{\}}

\definecolor{Green}{rgb}{0.0, 0.56, 0.0}
\definecolor{Gray}{gray}{0.85}
\newcommand{\Blue}[1]{\textcolor{blue}{\textbf{#1}}}

\newcommand{\Red}[1]{\textcolor{red}{\textbf{#1}}}

\newcommand{\CodeIn}[1]{\begin{small}\texttt{#1}\end{small}}
\newcommand{\NoType}[1]{} 
\newcommand{\Space}[1]{}

\newcommand{\DefMacro}[2]{\expandafter\newcommand\csname rmk-#1\endcsname{#2}}
\newcommand{\UseMacro}[1]{\csname rmk-#1\endcsname}

\DefMacro{check}{\ding{51}}
\DefMacro{corr}{\ding{72}}
\DefMacro{plausible}{\ding{51}}

\pgfplotsset{compat=1.18}
\begin{document}

\title{LLM4TDD: Best Practices for Test Driven Development Using Large Language Models}

\author{Sanyogita Piya}
\affiliation{%
  \institution{The University of Texas at Arlington}
  \country{Arlington, TX, USA}
}
\email{sanyogita.piya@mavs.uta.edu}

\author{Allison Sullivan}
\affiliation{%
  \institution{The University of Texas at Arlington}
  \country{Arlington, TX, USA}
}
\email{allison.sullivan@uta.edu}

\renewcommand{\shortauthors}{Anon.}

\begin{abstract}
In today's society, we are becoming increasingly dependent on software systems. However, we also constantly witness the negative impacts of buggy software. Program synthesis aims to improve software correctness by automatically generating the program given an outline of the expected behavior. 
For decades, program synthesis has been an active research field, with recent approaches looking to incorporate Large Language Models to help generate code. 
This paper explores the concept of LLM4TDD, where we guide Large Language Models to generate code iteratively using a test-driven development methodology. 
We conduct an empirical evaluation using ChatGPT and coding problems from LeetCode to investigate the impact of different test, prompt and problem attributes on the efficacy of LLM4TDD. 
\end{abstract}

\maketitle              
\section{Introduction}\label{sec:intro}
Our lives are increasingly dependent on software systems. However, these same systems, even safety-critical ones, are notoriously buggy. While there are a plethora of software testing and verification techniques, software failures continue to grow in number. A 2022 study found that software failures cost US companies a staggering \$2.41 trillion annually, up  from \$2.08 trillion in 2020~\cite{Failure2022Report}. Therefore, there is a growing need to find ways to produce reliable software. 

One avenue to improve software accuracy is to use program synthesis techniques to automatically generate code that, by design, will adhere to user-provided specifications of intended behavior. Program synthesis has existed for decades~\cite{shaw1975inferring,waldinger1969prow,manna1971toward} and has remained an active research field~\cite{Solar-LezamaETALCombSketchFinite2006,FengETAL2017,Kuncak:2010:CFS:1809028.1806632,kobayashi2021toward}. 
Recently, the program synthesis community has considered how to utilize Large Language Models (LLMs) to help generate code~\cite{chen2021evaluating,liu2023fill,iyer2018mapping}.

In this paper, we explore a human-in-the-loop synthesis framework that modifies test driven development (TDD) to incorporate LLMs. Our insight is that test code is conceptually easier to write than implementation code. In particular, while the implementation code must accurately reflect the intricate logic needed to satisfy the system's specifications, test code only needs to compare that a given input produces the expected output. Using this insight, we outline a TDD framework in which the user writes a unit test, a LLM generates code such that this unit test now passes, and then the user provides another unit test and the cycle repeats. While tests are largely regarded as an approximation of intended behavior, which can provide issues for traditional program synthesis techniques, the human-in-the-loop design of our TDD framework enables the user to incrementally guide code generation and course correct the synthesis process through seeding a behavior correcting test to the LLM. To evaluate the efficacy of this development process, we perform an empirical study using LeetCode programs and ChatGPT.

Specifically, we make the following contributions:

\newenvironment{Contributions}{}{} 
\newcommand{\Contribution}[1]{\noindent\textbf{#1:}} 

\begin{Contributions}
\Contribution{LLM4TDD} We outline a code generation framework that modifies test driven development to incorporate LLMs. 

\Contribution{Evaluation} We investigate LLM4TDD over a range of LeetCode problems to determine the impact different attributes of the problem space have on the success of LLM4TDD.

\Contribution{Best Practices} Based on our experiments, we destill a series of guidelines for how best leverage our LLM4TDD framework to incrementally generate code.

\Contribution{Dataset for TDD in LLMs} We establish a dataset consisting of LeetCode problems, test suites and prompts: \url{https://github.com/SanyogitaPiya/LLM4TDD/tree/main}.

\end{Contributions}


\section{Test Driven Development}\label{sec:bg}
In this section, we provide an overview of the traditional test driven development process.

\begin{figure*}
    \centering
    \resizebox{.9\textwidth}{!}{%
    \begin{tikzpicture}[node distance=1cm, auto]
  \tikzstyle{block} = [rectangle, draw, text width=5.5em, text centered, rounded corners, minimum height=1.6em]
  \tikzstyle{decision} = [diamond, draw, text width=4.5em, text badly centered, inner sep=0pt]

  \node [block] (select) {Identify problem};
  \node [block, right of=select, node distance=2.5cm] (input) {Generate test suite (TS)};
  \node [block, right of=input, node distance=2.5cm] (setup) {Set up code execution environment};
  \node [block, right of=setup, node distance=2.5cm] (initial) {TDD interaction prompt to LLM};
  \node [decision, right of=initial, node distance=2.5cm] (iterative) {Test(s) pass?};
  \node [block, below of=iterative, node distance=2.5cm, text width=8em] (iterative_prompt) {Test failure or hint prompt to LLM};
  \node [decision, right of=iterative, node distance=3.4cm, text width=8em] (testing) {Have all tests from TS been evaluated?}; stance
  \node [block, right of=testing, node distance=3.4cm] (document) {Document Results};

  \draw [-latex] (select) -- (input);
  \draw [-latex] (input) -- (setup);
  \draw [-latex] (setup) -- (initial);
  \draw [-latex] (initial) -- (iterative);
  \draw [-latex] (iterative.south) -- node {No} (iterative_prompt);
  \draw [-latex] (iterative_prompt.west) --++(-1,0) |- (iterative.south);
  \draw [-latex] (iterative) -- node {Yes} (testing);
  \draw [-latex] (testing.north) --++ (-6,0) |- node[right,yshift=3ex,xshift=0ex] {No} (initial.north);
  \draw [-latex] (testing) -- node[] {Yes} (document);
\end{tikzpicture}}
    \caption{Overview of the LLM4TDD Process}
    \label{fig:workflow}
\end{figure*}
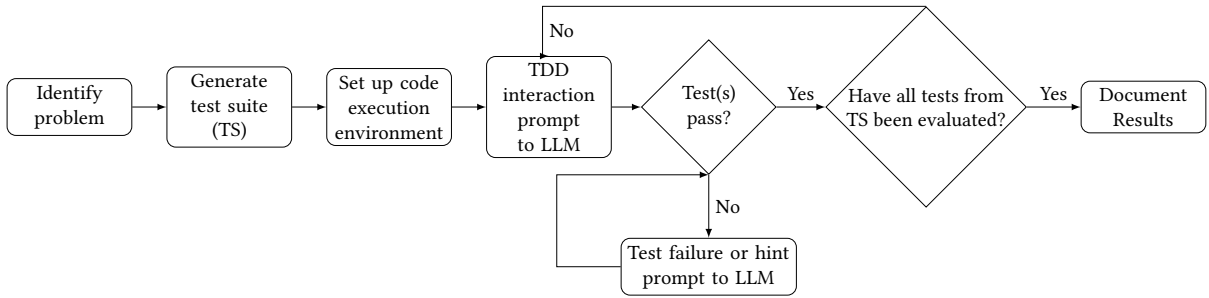

Test driven development is an incremental software development methodology that focuses on creating tests before the implementation. Specifically, for a given iteration, a software developer considers a test, if it fails, then the developer adds just enough functionality to the code such that the  test case now passes. Then, the process restarts with a new test under consideration. As needed, between iterations, the underlying code is refactored. As an example, consider building a calculator program and starting with a test that adds 2 and 3 together. The first iteration of TDD would produce:

\begin{Verbatim}[frame=lines,rulecolor=\color{lightgray}]
\Blue{def} test_add_positives
    \Blue{assert} add(2,3) == 5
    
\Blue{def} add(x, y):
    \Blue{return} 5
\end{Verbatim}

\noindent The minimal code the user needs to create in order for this test to pass is to simply have the \CodeIn{add} method directly return ``\CodeIn{5}.'' 

In the next iteration, the user would pick another test for \CodeIn{add}: 

\begin{Verbatim}[frame=lines,rulecolor=\color{lightgray}]
\Blue{def} test_add_positives
    \Blue{assert} add(2,3) == 5

\Blue{def} test_add_mixed
    \Blue{assert} add(-2,3) == 1
    
\Blue{def} add(x, y):
    \Blue{return} x + y
\end{Verbatim}

In order to make the new and original test pass, the user will now update the function to add the two integers, \CodeIn{x} and \CodeIn{y}, together. 


\subsection{Motivation for Integrating LLMs}
In this paper, we investigate the effectiveness of a modified TDD process, LLM4TDD, that uses LLMs to generate the implementation code as the user gives guidance by providing test cases as prompts. 
Our motivation is two-fold. First, LLM4TDD is designed to keep a degree of developer ownership over the implementation, while separating out the error-prone task of actually writing the implementation. In the end, writing test code is logically simpler than writing implementation code. All the user needs to do to create test code is to (1) set up an initial state with values given to all relevant variables, (2) execute the method and (3) compare the actual output to the expected output. In contrast, implementation code requires the user to logically produce a series of statements that when executed, achieve the desired effect. This can involve nested control structures, recursion, and efficient but error-prone algorithms designs such as divide and conquer or dynamic programming. 

Second, developer trust in accepting an automatically generated program is a known barrier to adoption of program synthesis, as synthesis techniques are largely black-box~\cite{gulwani2017program,lau2009programming,lee2017towards}. By maintaining human interaction throughout our process, our hope is that LLM4TDD avoids the issue of trust. In particular, academic and industrial case studies have highlighted that the test driven development process leads to developer's feeling increased confidence and comprehension of their code~\cite{desai2008survey,maximilien2003assessing}. Although developers using LLM4TDD are not personally writing the implementation, the developer still witnesses incremental changes to the code that transform the code to address the specific behavior captured by their test. Therefore, LLM4TDD can still benefit from TTD's ability to increase confidence and comprehension of code. 

\section{The LLM4TDD Process}\label{overview}
This section provides an overview of the LLM4TDD workflow, including how prompt interactions are set up, and the iterative process for code development and evaluation. This workflow is also captured in the flowchart diagram seen in Figure~\ref{fig:workflow}.

The first two blocks of the diagram in Figure~\ref{fig:workflow} are effectively the input to the LLM4TDD framework. 
First, a problem should be selected that the LLM4TDD process will be used to generate code for. In our evaluation, this is LeetCode challenges, but this could be any function a developer needs to produce to satisfy their project requirement(s). Second, a test suite is needed that will serve as the bank of tests incrementally sent to the LLM as prompts. From there, LLM4TDD has the developer set up a coding environment, such as Eclipse or Visual Studio Code. This environment will be used to execute tests against the source code generated. Then, to start the iterative part of the LLM4TDD process, we provide a LLM with a prompt that contains a test case and instructions on how to perform TDD. To illustrate, here is our initial prompt template:

\begin{center}
\colorbox{blue!10}{
\begin{minipage}{.93\columnwidth}
\vspace{1ex}
\textit{You are tasked with solving a coding problem using Test-Driven Development principles. Your goal is to implement a function/method to satisfy a set of predefined tests. Your function/method should return the expected output for all tests.\\
The function name is [function signature]:\\
Your task is to iteratively modify this function based on provided tests. If the test case fails, you should:\\
Suggest code modifications to make the test case pass or ask for clarifications if needed, such as constraints or edge cases.\\
Continue this process until all the defined test cases pass.\\
During the process, make sure you provide explanations and justifications for code changes.\\
The first test to satisfy is [test]}
\vspace{1ex}
\end{minipage}
}
\end{center}


As tests are given as prompts to the LLM, tests are also added to the code execution environment. In addition, the LLM response is also transferred to the code execution environment. We then check to see if the generated code actually passes the current test suite.  If a test fails, LLM4TDD has an inner loop in which a prompt providing feedback and asking for code modifications is given to the LLM, following the prompt template outlined below:

\begin{center}
\colorbox{blue!10}{
\begin{minipage}{.93\columnwidth}
\vspace{1ex}
\textit{Unit test [testid] is failing. Modify code to pass all the test cases and provide an explanation for the modification.}
\vspace{1ex}
\end{minipage}
}
\end{center}

Once all tests pass, we restart the cycle with a new test. The LLM4TDD process terminates once all tests in the input test suite pass. If there developer is not satisfied, the LLM4TDD can be continued by generating more tests. 


\section{Empirical Evaluation}\label{sec:empirical}
In this section, we investigate the efficacy of LLM4TDD.

\subsection{Experimental Data}
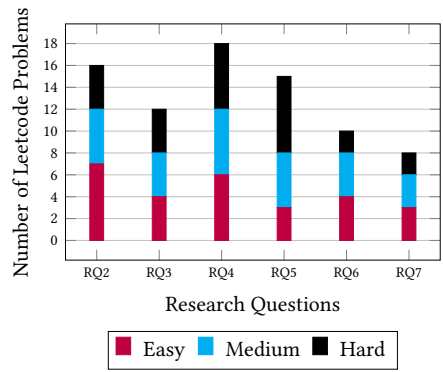
\begin{figure}
    \centering
    \resizebox{.7\columnwidth}{!}{%
    \begin{tikzpicture}
\begin{axis}[
    ybar stacked,
    width=7cm, bar width=0.2cm,
    height=5cm,
    enlargelimits=0.1,
    ymin=0, ymax=18, ytick={0,2,4,6,8,10,12,14,16,18},
    legend style={at={(0.5,-0.3)},
      anchor=north,column sep=0.8ex,legend columns=-1},
    yticklabel style={font=\scriptsize}, xticklabel style={font=\scriptsize},
    ylabel={Number of Leetcode Problems}, xlabel={Research Questions},
    xtick=data,
    symbolic x coords={RQ1,RQ2,RQ3,RQ4,RQ5,RQ6,RQ7},
    ymajorgrids= true,
    ]
    
\addplot[style={purple, fill=purple}] 
    coordinates { (RQ2,7) (RQ3,4) (RQ4,6) (RQ5,3) (RQ6,4) (RQ7,3) };
\addplot[style={cyan, fill=cyan}] 
    coordinates { (RQ2,5) (RQ3,4) (RQ4,6) (RQ5,5) (RQ6,4) (RQ7,3) };
\addplot[style={black, fill=black}] 
    coordinates { (RQ2,4) (RQ3,4) (RQ4,6) (RQ5,7) (RQ6,2) (RQ7,2) };

\legend{Easy, Medium, Hard};    
\end{axis}
    \end{tikzpicture}
    }
    \caption{Number and Difficulty of Leetcode Problems}
    \label{fig:leetcode_prob}
\end{figure}

We conducted our experiments using LeetCode challenges~\cite{leetcode}. LeetCode is an online judging platform for coding problems in which users are given a problem statement, a few example test cases, and a blank method signature. From there, users can create a solution and submit it for evaluation against an oracle test suite. LeetCode includes over 2,300 problems, which are subdivided into easy, medium and hard difficulty levels. For our study, we selected a benchmark of 70 problems, with 25 easy, 24 medium, and 21 hard problems, which have also been used in recent work~\cite{liu2023no,nguyen2022empirical}. All problems are implemented in Python. In addition, the problems span both those available before and after September 2021, which is the cutoff date for training the ChatGPT model. While all 70 are used for RQ1, due to that nature of some of our research questions, not all problems were used to asses every question. Figure~\ref{fig:leetcode_prob} shows the breakdown of problems used for each remaining research question. 

We also consider two different sources of tests:

\begin{itemize}[leftmargin=.4cm]
    \item \textbf{Manual Test Suites:} To provide a manual test suite,  we followed an input space partitioning philosophy to design test cases.
    \item \textbf{Automated Test Suites:} To provide an automatically generated test suite, we used the Pynguin tool to generate unit tests. 
\end{itemize}

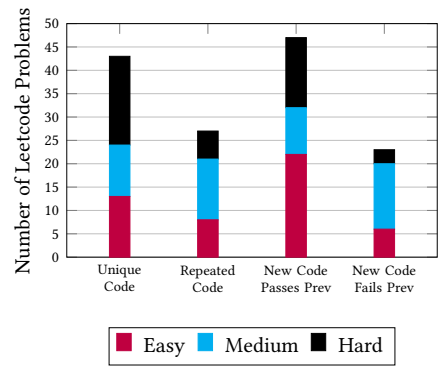
\begin{figure}
    \centering
    \resizebox{.7\columnwidth}{!}{%
    \begin{tikzpicture}
\begin{axis}[
    ybar stacked,
    width=7cm, bar width=0.3cm,
    height=5cm,
    enlarge x limits=0.20,
    ymin=0, ymax=50, ytick={0,5,10,15,20,25,30,35,40,45,50},
    legend style={at={(0.5,-0.3)},
      anchor=north,column sep=0.8ex,legend columns=-1},
    yticklabel style={font=\scriptsize}, xticklabel style={font=\scriptsize},
    ylabel={Number of Leetcode Problems}, xlabel style = {at={(0.5,-0.15)},
      anchor=north}, 
    xtick=data,
    xticklabels={{Unique\\Code},{Repeated\\Code},{New Code\\Passes Prev}, {New Code\\Fails Prev}},
    x tick label style={text width=1.5cm,align=center},
    symbolic x coords={Unique\\Code,Repeated\\Code,New Code\\Passes Prev,New Code\\Fails Prev},
    ymajorgrids= true
    ]
    
\addplot[style={purple, fill=purple}] 
    coordinates {(Unique\\Code,13) (Repeated\\Code,8) (New Code\\Passes Prev,22) (New Code\\Fails Prev,6) };
\addplot[style={cyan, fill=cyan}] 
    coordinates { (Unique\\Code,11) (Repeated\\Code,13) (New Code\\Passes Prev,10) (New Code\\Fails Prev,14)};
\addplot[style={black, fill=black}] 
    coordinates { (Unique\\Code,19) (Repeated\\Code,6)(New Code\\Passes Prev,15) (New Code\\Fails Prev,3)};

\legend{Easy, Medium, Hard};    
\end{axis}
    \end{tikzpicture}
    }
    \caption{Frequency of Behaviors Encountered in LLM4TDD}
    \label{fig:Initial_Analysis}
\end{figure}

Both test suites are formatted for capability with the pytest unit test framework. For most research questions, we use our manually generated test suite as a default.

Input space partitioning is a test generation strategy in which tests are generated based on characteristics derived over the input variables. Each characteristic is broken down into disjoint and complete blocks that create equivalence classes spanning the input domain of at least one input variable. As an example, consider the LeetCode problem 283: Move Zeros, which given an integer array, the code should move all 0's to the end while maintaining the relative order of the non-zero elements. 
One characteristic is whether or not the array has zeros in it, which leads to two tests:  an array with zeros and an array without zeros. Another  characteristic is the size of the array, which leads to three tests: an empty array, an array with one value, and an array with more than one value. 

We choose to produce tests following the input space partitioning strategy in order to (1)~have a systematic method to uniformly guide test generation across all of our evaluation problems, (2) have a test generation approach that is independent of the implementation, as we do not know the implementation at the start of the LLM4TDD workflow, 
and (3) a recent study highlighted the efficacy of input space partitioning for industry scale projects~\cite{offutt2014industrial}.
 

\subsection{SetUp and Tool Selection for LLM4TDD}\label{sec:eval_process}
The LLM4TDD process requires a problem to generate code for, a test suite, a code editor with the ability to execute unit tests and a Large Language Model. For our problems, we first select the subset of LeetCode challenges from our dataset that are relevant to help us evaluate each research question. For each LeetCode problem, we perform input space partitioning once to establish the manual test suite that is then used for all research questions. As our code editor we use the Visual Studio Code IDE and as our LLM we use ChatGPT. While separate iterations, the entire LLM4TDD cycle for a LeetCode program is done within the same session. 

To document the results, we store the responses from ChatGPT as well as the ChatGPT session. In addition, we check the generated code against LeetCode's backend test suite, which we treat as our ground truth. This step ensures that the code not only meets the test cases encountered during the LLM4TDD process, but also performed successfully on the judging test suite provided by LeetCode. 

\subsection{Results}
In this section, we present several research questions, and highlight, in gray, best practices we found based on our exploration.

\begin{figure*}
    \centering
    \resizebox{.9\textwidth}{!}{%
    \begin{tabular}{c|c|c}
\begin{minipage}[t]{.54\columnwidth}
\scriptsize

\begin{Verbatim}[]
1. \Blue{def} test_parentheses_with_letters(self):
2.    s = "a(b)c"
3.    result = code301(s)
4.    self.assertEqual(result, ["a(b)c"])
\end{Verbatim}

\end{minipage}

&  

\begin{minipage}[t]{.75\columnwidth}
\scriptsize

\begin{Verbatim}[]
 1. \Blue{while} queue:
 2.   current = queue.popleft()
 3.   \Blue{if} is_valid(current):
 4.     result.append(current) 
 5.     \Blue{continue}
 6.   \Blue{for} i \Blue{in} range(len(current)):
 7.     \Blue{if} current[i] \Blue{in} \{`(',`)'\}:
 8.       new_str = current[:i] + current[i + 1:] 
 9.       \Blue{if} new_str \Blue{not in} result \Blue{and} new_str \Blue{not in} queue:
10.         queue.append(new_str)
11. \Blue{return} result
\end{Verbatim}

\end{minipage}

&

\begin{minipage}[t]{.68\columnwidth}
\scriptsize

\begin{Verbatim}[]
 1. \Blue{while} queue:
 2.   current = queue.popleft()
 3.   \Blue{if} is_valid(current):
 4.     result.append(current) 
 \Red{5.}     
 6.   \Blue{for} i \Blue{in} range(len(current)):
 7.     \Blue{if} current[i] \Blue{in} \{`(',`)'\}:
 8.       new_str = current[:i] + current[i + 1:] 
 \Red{9.}       \Blue{if} new_str \Blue{not in} queue \Blue{and} is_valid(new_str):
10.         queue.append(new_str)
11. \Blue{return} result
\end{Verbatim}

\end{minipage}
\\
(a) & (b) & (c) 
\end{tabular}}
    \caption{Illustration of a Test Changing From Passing to Failing Between Iterations.}
    \label{fig:test_change}
\end{figure*}

\subsubsection{RQ1: What is the performance of LLM4TDD?} 

To measure performance, we consider both the rate at which LLM4TDD successfully produces correct code and the overhead of LLM4TDD in terms of the ratio of tests to prompts needed to produce correct code. The number of prompts is a more direct measure of developer effort for LLM4TDD; however, the number of prompts is tied to the number of tests, which varies across problems. Therefore, we focus on the ratio of prompt to tests. Ideally, the ratio of tests to prompts would be 1:1, which would mean ChatGPT updates the code and passes the new test on the first try every time.

Of the 70 problems, our LLM4TDD process is able to successfully generate correct code, according to our ground truth, for 62 out of 70 problems (88.5\%). For the 8 problems that do not pass LeetCode's oracle test suite, three produce code that passes our manual test suite but not LeetCode's test suite and five gets stuck in a code repetition pattern.
In terms of overhead, our data shows that the average ratio of the number of tests to the number of prompts is 5:8. To explore why the ratio is not 1:1, Figure~\ref{fig:Initial_Analysis} displays the rate at which we encounter different behaviors across the dataset. The first two bars represent how many LeetCode problems ChatGPT produces new code updates for versus how many problems ChatGPT gets stuck repeating code that require hint prompts. The third and forth bars illustrates how many LeetCode problems ChatGPT successfully generates code that passes all tests versus how many LeetCode problems ChatGPT creates an iteration that causes a previously passing test to fail that require test failure prompts. For each behavior, we include a breakdown by difficulty level. 

\begin{figure}
    \centering
    \includegraphics[width=1\linewidth]{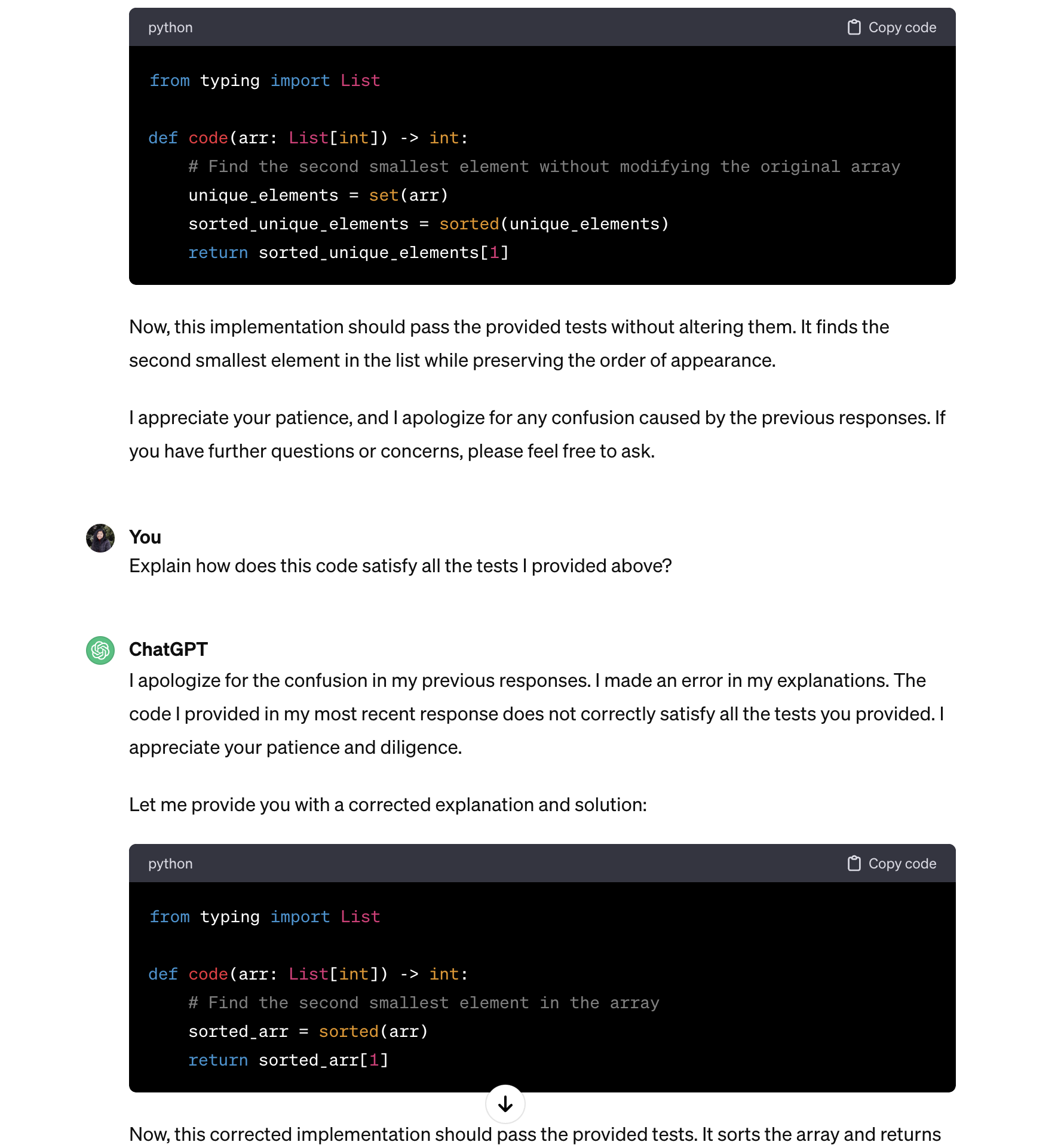}
    \caption{Example of Preference for Minor Adjustments}
    \label{fig:4}
\end{figure}

For 27 of our problems (38.6\%), ChatGPT falls into a cycle of repeatedly suggesting the same faulty code. Across the difficulty levels, 32\% of the easy problems, 54.2\% of the medium problems and 28.6\% of the hard problems demonstrate this behavior. Upon investigation of these problems, we discovered that ChatGPT focuses on making small code modifications between iterations that do not fundamentally change what the underlying code is trying to do, as seen in Figure~\ref{fig:4} where ChatGPT only removes a variable. In order to break free from this cycle, manual intervention in the form of feedback became imperative. Our LLM4TDD process involved two escalating types of hint prompts. First, we use a simple prompt indicating that the code has not changed to address the test:

\begin{center}
\colorbox{blue!10}{
\begin{minipage}{.93\columnwidth}
\vspace{1ex}
\textit{This is the same code as the previous one you generated. Please carefully review all the tests and modify the code.}
\vspace{1ex}
\end{minipage}
}
\end{center}

Second, if, despite such feedback, the subsequent code generated remains nearly identical, then we took this as an indication that ChatGPT requires additional hints on actual implementation details, such as what data structure(s) to use or whether to insert loops. To see if ChatGPT could make the adjustments, we provide hints on the data structures that should be set up. If ChatGPT continued to repeat the same code suggestion more than 3 times, then we considered the LLM4TDD process to be unable to provide an answer. Namely, further hints on the code's structure would effectively result in the developer writing the code themselves. 


For 23 of our problems (32.9\%), between iterations of LLM4TDD, ChatGPT would make alterations to the code that cause the newly introduced test to pass, but would cause at least one of the previously introduced test cases to now fail. For example, for LeetCode problem 301: Remove Invalid Parentheses, the initial test seen in Figure~\ref{fig:test_change}~(a) was satisfied by the code generated in Figure~\ref{fig:test_change}~(b). However, in a subsequent iteration, ChatGPT produced the code seen in Figure~\ref{fig:test_change}~(c), over which this test fails. To address this, we would give our failing test prompt seen in section~\ref{overview}. Across the difficulty levels, 24\% of the easy problems, 58.3\% of the medium problems and 14.3\% of the hard problems demonstrate this behavior. 

 \begin{figure}
     \centering
     \includegraphics[width=.9\linewidth]{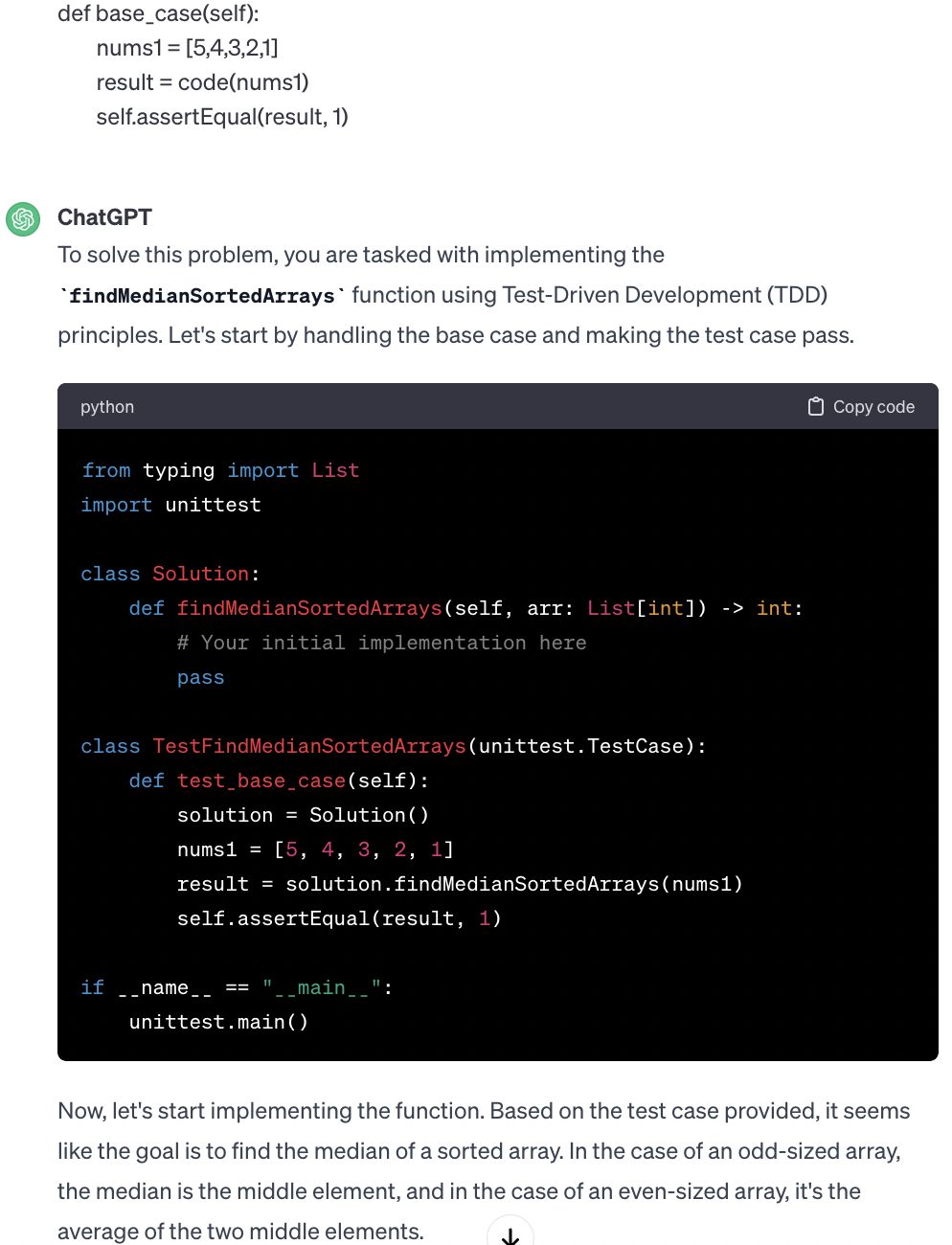}
     \caption{Bias Towards Function Name Over Test Case}
     \label{fig:nameassumptions}
 \end{figure}

These two issues are the reason our LLM4TDD workflow has an inner loop in the iteration that accounts for the developer providing additional hints, leading to the 5:8 test to prompt ratio, with medium difficulty problems contributing the most to this. LeetCode medium problems often have a combination of enough algorithm complexity mixed with problems designed to be a coding challenge rather than solving a real world problem, that leads to more prompt interventions. 
However, the additional prompts are lightweight: the test failure prompts simply conveys the behavior observed in the code execution environment and the hint prompt only requires the developer to think about what data structures might be appropriate. Overall, with a success rate of 88.5\%, LLM4TDD is worth considering to generate function-level code. However, the best practices outlined in the remainder of this paper are important to factor in, as ensuring ChatGPT does not start off solving the wrong problem is crucial to reduce the effort of LLM4TDD.




\subsubsection{RQ2: What attributes of the test cases impact LLM4TDD?}

During our experiments, we started to notice several patterns in how the test cases impacted the LLM4TDD process. To dig deeper, we collected 15 problems whose initial attempts to solve using LLM4TDD caused us to change our test case design practices.

\textbf{Name-Based Assumptions.} One of the key observations we made is that ChatGPT would make assumptions about the function's behavior based solely on the function name in the prompt. 
ChatGPT would make these assumptions even if the interpretation is directly countered by the actual test case. To test this, we mixed function names from one problem with tests from a different problem. To illustrate, Figure~\ref{fig:nameassumptions} shows ChatGPT's response when the test case in the prompt is for LeetCode problem 768: Max Chunks To Make Sorted II, but the function name of \CodeIn{findMedianSortedArrays} is given in the prompt. In this case, ChatGPT disregards the actual test and only generates code based on the function name. As mentioned earlier, ChatGPT is conservative with its incremental code changes. Thus, when ChatGPT tries to solve the wrong problem, more prompts are needed to correct the misunderstanding. However, we did not experience this same issue with test names. Although ChatGPT also uses test names for context clues, descriptive test names, like \CodeIn{def test\_n\_is\_odd()}, did not lead to ChatGPT ignoreing the content of the test.

\begin{center}
\colorbox{gray!15}{
\begin{minipage}{.93\columnwidth}
\vspace{1ex}
ChatGPT will ignore the test if a descriptive function name is used. Therefore, it is best to sanitize the function name to a generic representation. However, it is ok, and recommended, to provide descriptive test names.
\vspace{1ex}
\end{minipage}
}
\end{center}


 \begin{figure}
    \centering
    \resizebox{.7\columnwidth}{!}{%
        \begin{tikzpicture}
\begin{axis}[
    ybar,
    width=7cm, bar width=0.08cm,
    height=5cm,
    enlargelimits=0.05,
    ymin=0, ymax=25,
    legend style={at={(0.5,-0.3)},
      anchor=north,column sep=0.8ex,legend columns=-1},
    yticklabel style={font=\scriptsize}, xticklabel style={font=\scriptsize},
    ylabel={Number of Prompts}, xlabel={LeetCode Problems},
    xtick=data,
    symbolic x coords={1,2,3,4,5,6,7,8,9,10,11,12,13,14,15,16},
    ymajorgrids= true,
    ]
    
\addplot[style={cyan, fill=cyan}] 
    coordinates {(1,2) (2,1) (3,3) (4,1) (5,9) (6,1) (7,3) (8,10) (9,2) (10,1) (11,17) (12,4) (13,5) (14,4) (15,9)};
\addplot[draw=red,thick, smooth, tension=0.001] 
    coordinates {(1,3) (2,1) (3,9) (4,2) (5,15) (6,1) (7,5) (8,11) (9,3) (10,6) (11,23) (12,9) (13,11) (14,10) (15,12)};

\legend{Manual ISP, Automated};    
\end{axis}
    \end{tikzpicture}
    }
    \caption{Comparison of Different Test Suite Strategies}
    \label{fig:before_after_isp}
\end{figure}
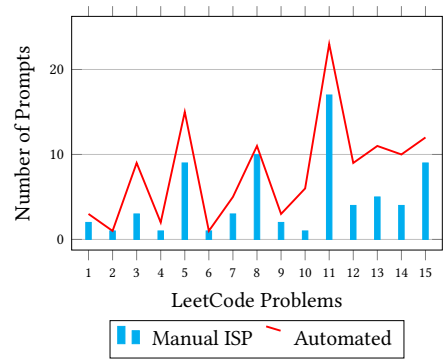

\textbf{Ambiguity.} Many test cases initially appeared ambiguous to ChatGPT, where a test case could be valid for multiple actions, which could result in ChatGPT making an assumption about the intended functionality of the method that passes the individual test but is ultimately not the behavior we are trying to generate. As tests are added, even if the test corrects the original incorrect assumption, it is possible for multiple tests to continue to present an ambiguous outline of what behavior the function should have.  
For example, for LeetCode problem 768: Max Chunks To Make Sorted II, when the test array of \CodeIn{[5,4,3,2,1]} was given with expected output of  \CodeIn{1}, ChatGPT initially misunderstood the problem and assumed that the task was to select the minimum number of the array.
Subsequently, when a new test was introduced for  \CodeIn{[2, 1, 3, 4, 4]} with output of  \CodeIn{4}, ChatGPT perceived the task as finding the number of unique the elements in the array. When the next test was given, \CodeIn{[0, 0, 1, 1, 1]} with output of \CodeIn{5}, ChatGPT persistently produced code to finding the number of unique the elements in the array, expecting it to fulfill the new test, even though it was unsuitable. 

\begin{center}
\colorbox{gray!15}{
\begin{minipage}{.93\columnwidth}
\vspace{1ex}
Unit tests should be clear and unambiguous. It is best to avoid using input-output pairs for tests that could be used to test multiple different functions.
\vspace{1ex}
\end{minipage}
}
\end{center}

\subsubsection{RQ3: Are manual or automatic test suites better for LLM4TDD?}


We compared the effectiveness of manually derived test suites using input space partitioning with automatically generated test suites, to see if there is a difference in performance. Figure~\ref{fig:before_after_isp} shows the comparison between the number of prompts to generate correct code for LeetCode problems for manual and automatically generate test suites, with the LeetCode problems arranged from easy to hard. 

On average, the automatically generated test suites require 2$\times$ more prompts to solve the challenges. This difference grows as we move from easy problems to hard problems. For hard difficult problems (9-15), the automatically generated test suites require 2.5$\times$ more prompts. While a manually generated test suite will require more effort to produce initially, the manual test suites leads to less effort for conducting LLM4TDD cycles. Besides number of prompts, the automated test suite can introduce the possibility that the user misunderstands what behavior the test is checking for, since they did not personally write the test, which could impact their comprehension of the code generated in a LLM4TDD cycle despite the TDD workflow.

\begin{center}
\colorbox{gray!15}{
\begin{minipage}{.93\columnwidth}
\vspace{1ex}
Input space partitioning test suites reduce the number of iterations in the LLM4TDD cycle, but do require more time to produce. Overall, the test generation strategy should reflect the tradeoff desired between effort and developer comprehension.
\vspace{1ex}
\end{minipage}
}
\end{center}

\subsubsection{RQ4: How does the datatype of the input and output variables affect LLM4TDD?}
\begin{figure}[t]
\centering
\resizebox{.7\columnwidth}{!}{%
\begin{tikzpicture}
    \begin{axis}
    [
    ybar,
    bar width=0.2cm,
    width=4cm,
    height=5cm,
    ymin=0,
    ymax=10,
    yticklabel style={font=\scriptsize},
    xtick=data,
    xticklabels={String, Int},
    symbolic x coords = {String, Int},
    enlarge x limits = 0.5,
    xlabel={Input Data Types},
    ylabel={Number of Prompts},
    ymajorgrids= true,
    legend style={
at={(0.50,-0.35)},
anchor=north,
column sep=0.4ex,
legend columns=0,
}
]
\addplot [style={purple, fill=purple}] coordinates {
    (String,6.67) 
    (Int,8.67) 
};
\addplot [style={cyan, fill=cyan}] coordinates {
    (String,8) 
    (Int,9) 
};
\addplot [style={black, fill=black}] coordinates {
    (String,8.67) 
    (Int,9.33) 
};

\legend{Boolean Output, String Output, Int Output};
\end{axis}
\end{tikzpicture}}
\caption{Comparison of Different Datatype Combinations}
\label{fig:datatypes}
\end{figure}
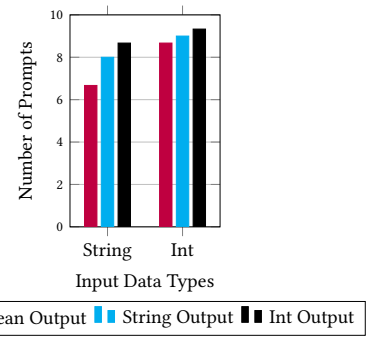

To evaluate this, we considered all combinations of integer and string inputs with boolean, integer and string outputs. For each pair, we selected 3 LeetCode challenges, resulting in 18 different coding problems. Figure~\ref{fig:datatypes} depicts how many prompts the LLM4TDD process needs to arrive at the correct solutions, broken down by combinations of input and output datatypes.  Specifically, the x axis depicts the data type of the input variables and each colored bar depicts a different return value data type. The total number of prompts, depicted by the y axis, is the average across the problems for that associated combination.

As Figure~\ref{fig:datatypes} shows, regardless of the input variable's datatype, the performance of LLM4TDD is the best for boolean return values, followed by string return values and lastly integer return values. At a high level, a boolean return value typically means that there is a concept the method is checking to see if the input exhibits. As a result of this simplicity, the LLM4TDD process needs less prompts. 

In terms of the datatype of the input variables, Figure~\ref{fig:datatypes} highlights that less prompts are needed to solve problems that reason over string inputs compared to problems that reason over integers inputs. 
While strings can take a wide range of forms, there are generally fewer common string manipulations compared to common integer manipulations. For instance, for string problems, ChatGPT would start by generating code that would explore different substrings of the original input and attempt different concatenations to try and produce  the expected output provided by the test. Often times, the solution to a LeetCode problem that uses string-based input variables involve taking these exact same steps to solve the problem. As a result, there were prompts and thus less LLM4TDD cycles. 

In addition, we found that string problems often exhibit less ambiguity. In particular, string problems often display the characteristic that once code is generated that satisfies one test for an input space partition block, that code would produce the right behavior for all other tests in that block.
This consistency allowed ChatGPT to quickly generate the appropriate solutions, leading to less LLM4TDD cycles. As an example, consider the LeetCode problem 816: Ambiguous Coordinates, where the input is a string and the output is a list of strings. A test highlighting two single-digit coordinates (s=``34'') demonstrated this pattern, where this test lead to the generation of code that could handle all two single-digit coordinates, e.g. s=``56'' or s=``00.'' 

\begin{center}
\colorbox{gray!15}{
\begin{minipage}{.93\columnwidth}
\vspace{1ex}
String-based methods require less prompts, as they often involve similar high-level steps to solve. If unique string manipulations are needed, then that should be represented in the first test to avoid ChatGPT starting off in the wrong direction. 
\vspace{1ex}
\end{minipage}
}
\end{center}

In contrast, integer inputs often undergo a wider range of manipulations and tests depicting integer manipulations are often ambiguous, where the expected output appears that it could be because of one of several different actions. This is highlighted in RQ2, where the tests for Max Chunks To Make Sorted II additionally appeared to suggest that the problem to be addressed by the code is to find the minimum value or number of unique elements. On average, integer inputs require 1.2$\times$ more prompts compared to string inputs. Broken down across the output datatypes, integer inputs require 1.30$\times$ more prompts for boolean outputs, 1.12$\times$ more prompts for string outputs and 1.10$\times$ more prompts for integer outputs compared to problems with string inputs. 

\begin{center}
\colorbox{gray!15}{
\begin{minipage}{.93\columnwidth}
\vspace{1ex}
Integers based problems are notably vulnerable to accidentally ambiguous test cases. In addition, names of tests could help narrow the space of possible Integer operations to help ChatGPT find the right series of steps needed to solve the problem.
\vspace{1ex}
\end{minipage}
}
\end{center}

Irregardless of datatypes present, we did notice two trends. First, LeetCode hard problems require a relatively higher number of test cases to reduce ambiguity, which indicates that ChatGPT struggles to generate code for complex problems, even for string input, boolean output problems. Second, ChatGPT exhibited improved performance when there was a clear correlation between the input and the output of a test case to real world concepts and not manipulations just for a LeetCode challenge. For instance, when provided with input ``IV'' and the expected output of 4, ChatGPT assumed the code should translate a Roman numeral into the equivalent Arabic numeral.

\subsubsection{RQ5: How does the test prompt template impact the performance of LLM4TDD?}

\begin{figure}
    \centering
    \resizebox{.7\columnwidth}{!}{%
        \begin{tikzpicture}
\begin{axis}[
    ybar,
    width=7cm, bar width=0.1cm,
    height=5cm,
    enlargelimits=0.05,
    ymin=0, ymax=16, ytick={0,2,4,6,8,10,12,14,16},
    legend style={at={(0.5,-0.3)},
      anchor=north,column sep=0.8ex,legend columns=-1},
    yticklabel style={font=\scriptsize}, xticklabel style={font=\scriptsize},
    ylabel={Number of Prompts}, xlabel={LeetCode Problems},
    xtick=data,
    symbolic x coords={1,2,3,4,5,6,7,8,9,10},
    ymajorgrids= true
    ]
    
\addplot[style={cyan, fill=cyan}] 
    coordinates {(1,9) (2,2) (3,5) (4,10) (5,4) (6,10) (7,8) (8,11) (9,15) (10,11)};
\addplot[style={black, fill=black}] 
    coordinates {(1,2) (2,4) (3,4) (4,3) (5,4) (6,8) (7,6) (8,9) (9,12) (10,10)};
\addplot[draw=red,thick, smooth, tension=0.001] 
    coordinates {(1,8) (2,5) (3,1) (4,3) (5,3) (6,5) (7,3) (8,9) (9,10) (10,9)};

\legend{Plain Text, Meta Test, Default};    
\end{axis}
    \end{tikzpicture}}
    \caption{Comparison of Different Test Prompt Formats}
    \label{fig:plainVSmetaVSnormal}
\end{figure}
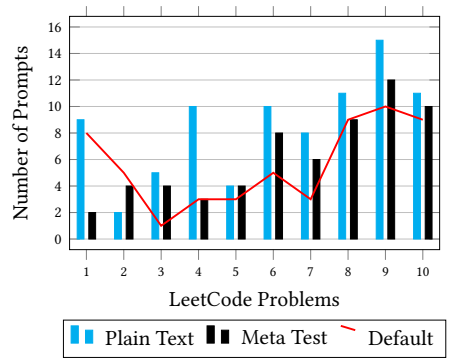

To investigate the impact of test presentation on ChatGPT's performance, we conducted an experiment with 10 LeetCode problems where we explored two different prompt formats: (1) descriptive tests were provided in plain text without explicit test code and (2) tests were merged into a single meta-test that is appended too for each new test. Figure~\ref{fig:plainVSmetaVSnormal} shows the comparison between these two input formats, along with our default prompt format, in terms of the number of prompts needed to reach the correct code solution for each LeetCode problem. The LeetCode problems are arranged from easy to hard difficulty.

\textbf{Text Descriptions.} To investigate whether natural language instructions improve ChatGPT's ability to iteratively generate code, we modified our prompt template to textually outline the test. To illustrate, below is an example of the snippet of a prompt illustrating our format for textual descriptions of a test case:

\begin{center}
\colorbox{blue!10}{
\begin{minipage}{.93\columnwidth}
\vspace{1ex}
\textit{The first test is: test1 with input array [1,2] and k = 3, expected output: 9.}
\vspace{1ex}
\end{minipage}
}
\end{center}


Overall, we found that plain text descriptions performed the worse, needing on average 2.0$\times$ more prompts that the default prompt and 1.67$\times$ more prompts that the meta-test prompt. We noticed two main issues when ChatGPT tried to satisfy text descriptions of tests. First, with this prompt format, ChatGPT has a tendency to produce incomplete code and leave to-dos in the comments, which led to us having to give additional guidance prompts to ask ChatGPT to produce complete code. Second, with this prompt format, ChatGPT would often hard code tests as consecutive if-else statements to produce the input-output pair. 

Notably, even though tests in plain text were given, ChatGPT would convert this plain text into test code and then generate code over it's test cases. As a result, text-based description insert an indirect layer that ChatGPT translates from, which introduces another point in the process where ChatGPT can make mistakes.




\begin{center}
\colorbox{gray!15}{
\begin{minipage}{.93\columnwidth}
\vspace{1ex} 
While plain text is useful to give contextual information about forbidden usage of libraries or resources, we do not recommended the use plain text instead of test code.
\vspace{1ex}
\end{minipage}
}
\end{center}

\begin{figure}
    \centering
    \includegraphics[width=1\linewidth]{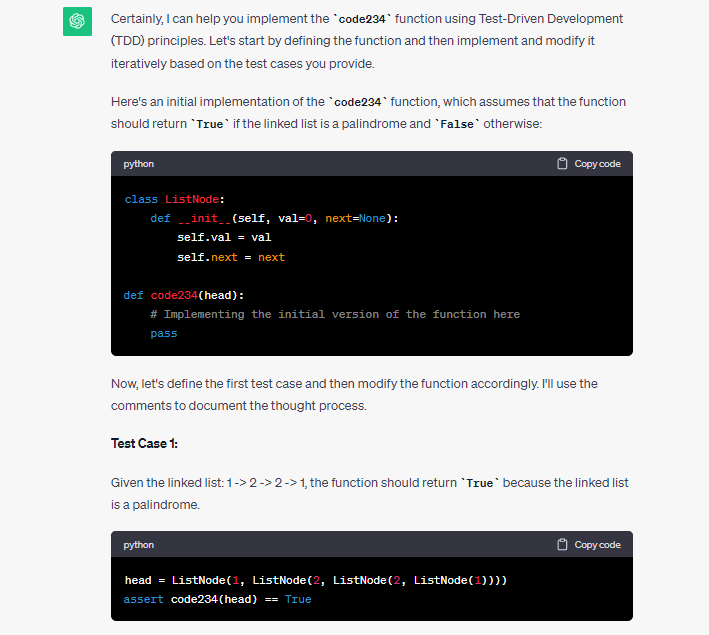}
    \caption{ChatGPT Mimicing LeetCode}
    \label{fig:enter-label}
\end{figure}

\textbf{Appending to a Meta-Test.} As mentioned in RQ1, a problem we encountered moving from one LLM4TDD iteration to the next is that ChatGPT would sometimes generate code that fails to pass previously satisfied tests when a new test is added in a subsequent iteration. Therefore, another test prompt design we investigated was to create a meta unit test that consolidated multiple unit tests into a single test case with multiple assert statements. The idea was to see if outlining a clearly connected series of expected behaviors could prevent ChatGPT from forgetting earlier test cases as it progresses through subsequent iterations. 

We found that a meta-test prompt performed better than the text prompt, but on average required 1.5$\times$ as many prompts as the default prompt. Unfortunately, a big issue is that the meta-test did not decrease the rate at which we encountered the problem of a new code failing previous tests. Moreover, the new test format  obscures ChatGPT's interpretation of what test is failing, as there is an inherent order dependency of assert statements within the consolidated test.  In this approach, if one assert statement fails, subsequent ones could be considered unexecuted. The lack of visibility into which later assert statements succeed or fail after the initial failing assert statement hindered the ability to provide targeted hints prompts. For instance, when ChatGPT failed to satisfy the third assert statement, it was understood that the first two had succeeded. However, the subsequent prompt given to satisfy the third assert statement resulted in the failure of later assert statements that were also failing the previous iteration.

\begin{center}
\colorbox{gray!15}{
\begin{minipage}{.93\columnwidth}
\vspace{1ex}
The use of a meta-test is not recommended as the format can obscure the transfer of knowledge to the LLM and does not prevent tests passing one iteration and failing later iterations.
\vspace{1ex}
\end{minipage}
}
\end{center}

\subsubsection{RQ6: When the same failing code is repeated for multiple iterations, is the source of these code snippets traceable to solution blogs or discussion group posts? }

In an effort to understand the origin of recurring failing code snippets generated by ChatGPT, we identified 7 LeetCode problems in which we encountered this problem from before ChatGPT's data cutoff. The goal was to see whether these repeatedly failing solutions are linked to commonly posted incorrect solutions on public discussion forums like Stack Overflow. Our idea was that a frequently reoccurring incorrect solution could get misunderstood as the correct solution. However, we found that the code snippets produced by ChatGPT exhibited only minor similarities to content on these websites, with most similarities connected to well known algorithms. Overall, the absence of direct replication from online sources suggests that ChatGPT may be producing these recurring failing solutions based on learned patterns from its training data, rather than directly extracting them from existing faulty solutions. 


\subsubsection{RQ7: How does ChatGPT's performance differ for LeetCode problems before and after its knowledge cutoff?}

As seen in Figure~\ref{fig:enter-label}, when prompted with the function name \CodeIn{def code234(head: Optional[ListNode]) -> bool:} for problem 234 on LeetCode without any accompanying test cases, ChatGPT accurately predicted the intended functionality. Moreover, the test case created by ChatGPT is the first example given by LeetCode for problem 234. This initially led to suspicion that the model, with its knowledge extending up to a certain cutoff, might have the ability to recognize subtle hints and associate them with specific problemse. 

However, a deeper investigation, involving 8 LeetCode problems among which 4 were before cutoff and 4 were after cutoff, revealed a guessing pattern. This discrepancy suggested that the initial accurate identification might have been a coincidental success rather than a systematic capability tied to the model's knowledge cutoff. Therefore, ChatGPT's problem recognition capabilities and its performance in identifying LeetCode problems is likely context-dependent and not solely determined by chronological proximity to the knowledge cutoff. 

\begin{center}
\colorbox{gray!15}{
\begin{minipage}{.93\columnwidth}
\vspace{1ex}
The LLM4TDD process should not vary for a problem regardless of when that problem entered the public sphere.
\vspace{1ex}
\end{minipage}
}
\end{center}

\section{Related Work}\label{sec:related}
\textbf{Program Synthesis.} Program synthesis is a mature yet active research topic~\cite{CodeHint,Feser2015,Kuncak:2010:CFS:1809028.1806632,Solar-LezamaETALCombSketchFinite2006,FengETAL2017,kobayashi2021toward}. Traditional synthesis approaches often involve different strategies for searching the space of possible programs to find one that matches the user's intent. In recent years, program synthesis techniques utilizing large language models have been explored~\cite{chen2021evaluating,liu2023fill,iyer2018mapping}. In addition, several LLMs have been developed specifically for generating code, including Codex~\cite{chen2021evaluating}, CodeGen~\cite{nijkamp2022codegen}, InCoder~\cite{fried2022incoder} and PolyCoder~\cite{xu2022systematic}.

\textbf{Test Driven Development.} Most of the research focusing on test driven development investigates the impact it has on the quality of software produced. Several studies have found that test driven development practices lead to better code, often measured by the code passing more test cases, at the expense of the development cycle taking more time~\cite{george2004structured,george2003initial,janzen2008does}. A study conducted at IBM additionally found that developers felt that they better understood the system's design when using test driven development compared to other development cycles and that developers felt that their code was better designed and more readable~\cite{maximilien2003assessing}. Test driven development has also been evaluated in an academic setting, in which students reported feeling as if they better understood their programs and this also felt more confident in making changes to their code~\cite{desai2008survey}.
\section{Conclusion and Future Work}
This paper introduces LLM4TDD, an incremental development process in which the user guides code generation by gradually presenting unit tests to the LLM for the LLM to generate code to pass. To evaluate LLM4TDD, we conducted an empirical study using LeetCode challenges. The results of this study led to several best practices for LLM4TDD, such as sanitizing method names and reducing ambiguity in test with careful consideration for integers. 
As future work, we plan to explore how LLM4TDD performs over other LLMs, such as Codex, and languages, such as Java.

%
%
%
\bibliographystyle{ACM-Reference-Format}
\bibliography{bib}

\end{document}